\documentstyle[12pt]{article}
\begin{document}
\hfill{NCKU-HEP-98-10}\par
\vskip 1.0cm

\centerline{\large\bf The $\Lambda_b\to pl{\bar\nu}$ decay in
perturbative QCD}
\vskip 1.0cm
\centerline{Hsien-Hung Shih and Shih-Chang Lee}
\vskip 0.5cm
\centerline{Institute of Physics, Academia Sinica,}\par
\centerline{Taipei, Taiwan 115, Republic of China}
\vskip 0.5cm
\centerline{Hsiang-nan Li}
\vskip 0.5cm
\centerline{Department of Physics, National Cheng-Kung University,}\par 
\centerline{Tainan, Taiwan 701, Republic of China}
\vskip 1.0cm

\centerline{\bf abstract}
\vskip 0.5cm

We develop perturbative QCD factorization theorems for the exclusive
semileptonic heavy baryon decay $\Lambda_b \to pl{\bar\nu}$, in which the
relevant hadronic form factor is expressed as the convolution of a hard
subamplitude with the $\Lambda_b$ baryon wave function and the proton wave
function. The specific evolution scale for the proton wave function,
determined from the best fit to the data of the proton form factor, is
adopted. The Sudakov resummation for a heavy-light system (the $\Lambda_b$
baryon) and the quark-level decay diagrams with at least one hard gluon
attaching the $b$ quark, which were ignored in the literature, are
incorporated. It turns out that these additional ingredients are important:
the neglect of the former and the latter reduces the results of the
form factor by factors 1.5 and 3, respectively. We present the
predictions of the form factor and of the proton energy spectrum for the
various choices of the parameter involved in the $\Lambda_b$ baryon wave
function.

\newpage

\centerline{\large \bf I. INTRODUCTION}
\vskip 0.5cm

Recently, we have applied perturbative QCD (PQCD) factorization theorems
to various $B$ meson decays, and progresses have been made [1-8]. For
exclusive processes \cite{LY1,YL}, the form factors involved in decay
spectra and decay rates are factorized into the convolution of a hard
subamplitude with meson wave functions. The former is process-dependent
and calculable in perturbation theory at the parton level. The latter,
absorbing long-distance dynamics of $B$ meson decays, are universal
(process-independent), and must be extracted from experimental data or
derived by using nonperturbative methods. For example, we have determined
the $B$ meson wave function from the best fit to the experimental data of
the branching ratio ${\cal B}(B\to K^*\gamma)$ \cite{Lk}.
The resummation \cite{CS,BS} of double logarithms from the overlap of
collinear and soft enhancements in radiative corrections to meson wave
functions plays an important role in the analyses. The resultant Sudakov
factor suppresses the long-distance contributions, and improves the
applicability of PQCD around the few GeV scale \cite{LS}.

The Wilson evolution in effective field theory has been incorporated into
the above PQCD formalism, leading to the three-scale factorization theorem
for nonleptonic decays \cite{YL}. As the evolution scale runs to below the
$b$ quark mass $M_b$ and the $c$ quark mass $M_c$, the constructive and
destructive interferences between the external and internal $W$-emission
contributions involved in bottom and charm decays, respectively, appear.
Nonfactorizable and nonspectator contributions can be evaluated
systemmatically, which make possible the simultaneous explanation of the
ratios $R={\cal B}(B\to J/\psi K^*)/{\cal B}(B\to J/\psi K)$ and
$R_L={\cal B}(B\to J/\psi K_L^*)/{\cal B}(B\to J/\psi K^*)$
in the $B\to J/\psi K^{(*)}$ decays \cite{GKP}. With the further inclusion
of the all-order soft gluon exchanges, the mechanism of the opposite signs
of nonfactorizable contributions in bottom and charm decays has been
understood \cite{LT}.

Similarly, inclusive decay spectra and decay rates of the $B$ meson are
factorized into the convolution of a hard $b$ quark decay subamplitude with
a $B$ meson distribution function and several jet functions for light
energetic final states \cite{LY2,CC}. The distribution function, being the
outcome of the resummation of nonperturbative power corrections to the
processes, has been extracted from the photon energy spectrum of the decay
$B\to X_s\gamma$ \cite{LY3}. The jet functions collect the double logarithms
appearing at the end points of decay spectra, whose resummation gives the
Sudakov factor. The three-scale factorization theorem also applies to
inclusive nonleptonic decays. Using the above formalism, we have observed
that the single-charm\ mode $b\to c{\bar u}d$ is enhanced more than the
double-charm mode $b\to c{\bar c}s$ is. The semileptonic branching ratio
$B_{\rm SL}\equiv{\cal B}(B\to Xl{\bar\nu})$ is then reduced without
increasing the charm yield $n_c$ per $B$ decay, and the large $B_{\rm SL}$
controversy in the conventional approach based on heavy quark effective
theory is resolved \cite{CC}. Our predictions for the absolute lifetimes of
the $B$ meson and of the $\Lambda_b$ baryon are also consistent with the
data, such that the lifetime ratio $\tau(\Lambda_b)/\tau(B_d)$ can be
explained \cite{CLY}.

With the above successes, it is natural to extend the PQCD formalism to
more complicated heavy baryon decays. In this paper we shall start with
the simplest exclusive mode $\Lambda_b\to pl\bar\nu$, taking into account
the Sudakov resummation for a heavy-light system (the $\Lambda_b$ baryon in
this case), investigating the applicability of PQCD to heavy baryon decays,
and understanding the sensitivity of predictions to the variation of the
$\Lambda_b$ baryon wave function, which remains unknown. It is expected
that a PQCD analysis of heavy baryon decays is not as reliable as that of
heavy meson decays, since partons in the former case are softer and Sudakov
suppression is weaker. We shall show that PQCD for the decay
$\Lambda_b\to pl{\bar \nu}$ is reliable only at the high end of the proton
energy. Therefore, our approach may be appropriate for exclusive nonleptonic
heavy baryon decays, which will be studied in the future.

The first attempt to apply the PQCD factorization theorem to the
$\Lambda_b\to p l \bar\nu$ decay was made by Loinaz and Akhoury \cite{RA}.
However, they did not consider the Sudakov resummation for the $\Lambda_b$
baryon, and their evaluation of the hard subamplitude was not complete:
the quark-level decay diagrams with at least one exchanged gluon attaching
the $b$ quark were ignored. Our analysis shows that these two
ingredients are in fact important. The neglect of the former and the
latter reduces the results of the relevant form factor by factors 1.5 and 3,
respectively. We shall adopt the King-Sachrajda (KS) model \cite{KS} for the
proton wave function, along with the specific choice of its evolution scale,
which were determined from the best fit to the experimental data of the
proton form factor \cite{L,KLS}. Furthermore, the full expression of the
Sudakov factor with the accuracy up to next-to-leading logarithms \cite{YL}
will be inserted.

In Sect. II we briefly explain how to factorize the perturbative and
nonperturbative contributions to the decay $\Lambda_b\to pl\bar\nu$
into the hard subamplitude and the baryon wave functions, respectively. The
large logarithms contained in these convolution factors are organized
in Sect. III. The factorization formula for the relevant form factor
is presented in Sect. IV. Numerical analyses are performed in Sect. V,
where the behaviors of the form factor and of the proton energy 
spectrum are exhibited. Section VI is the conclusion.

\vskip 1.0cm

\centerline{\large\bf II. FACTORIZATION THEOREMS}
\vskip 0.5cm

The amplitude for the semileptonic decay $\Lambda_b\to pl{\bar\nu}$ is
written as
\begin{eqnarray}
{\cal M} &=& \frac{G_F}{\sqrt{2}}V_{ub}{\cal M}_{\mu}
 [ {\bar l}\gamma^{\mu} (1-\gamma_5)\nu_l ],
\end{eqnarray}
where $G_F$ is the Fermi coupling constant, and $V_{ub}$ the
Cabibbo-Kobayashi-Maskawa (CKM) matrix element. All QCD dynamics is
contained in the hadronic matrix element,
\begin{eqnarray}
{\cal M}_{\mu} &=& \langle P(p^\prime)|\bar{u}\gamma_{\mu} (1-\gamma_5)b|
\Lambda_b(p)\rangle\;,
\end{eqnarray}
where $p$ ($p^\prime$) is the momentum of the $\Lambda_b$ baryon (proton).
The general structure of a matrix element of charged weak currents
between baryon states is expressed as
\begin{eqnarray}
\langle B^\prime,p^\prime, s^\prime | j^\mu |B, p, s \rangle &=&
\bar{B}'(p^\prime, s^\prime)[ f_1(q^2)\gamma^\mu
-if_2(q^2)\sigma^{\mu\nu}q_\nu
\nonumber\\
& &\hspace{1.0cm}+f_3(q^2)q^\mu ] B(p,s)\;,
\nonumber \\
\langle B^\prime,p^\prime, s^\prime | j_5^\mu |B, p, s \rangle &=&
\bar{B}'(p^\prime, s^\prime)[ g_1(q^2)\gamma^\mu\gamma_5
-ig_2(q^2)\sigma^{\mu\nu}\gamma_5q_\nu
\nonumber\\
& &\hspace{1.0cm}+g_3(q^2)\gamma_5q^\mu ] B(p,s)\;,
\end{eqnarray}
where $j^\mu$ is the vector current, $j_5^\mu$ the axial vector current,
$B(p,s)$ the spinor of the initial-state baryon $|B, p, s \rangle$ with $p$
and $s$ the momentum and spin, respectively. The spinor
${\bar B}'(p^\prime, s^\prime)$ is associated with the final-state baryon
$\langle B^\prime,p^\prime, s^\prime |$. The form factors $f_i(q^2)$ and
$g_i(q^2)$ depend only on the momentum transfer $q^2=(p-p^\prime)^2$.

In the case of massless leptons with
\begin{eqnarray}
q_\mu \bar{l}\gamma^\mu(1-\gamma_5)\nu_l= 0 \;,
\end{eqnarray}
it is easy to observe that only the four form factors $f_1$, $f_2$, $g_1$,
and $g_2$ contribute. We rewrite the hadronic matrix element as
\begin{eqnarray}
{\cal M}_\mu &=&\bar{P}(p^\prime) [L_1(q^2)\gamma_\mu(1-\gamma_5)+
R_1(q^2)\gamma_\mu(1+\gamma_5)
\nonumber\\
& &+L_2(q^2)i\sigma_{\mu\nu}(1-\gamma_5)q^\nu
+R_2(q^2)i\sigma_{\mu\nu}(1+\gamma_5)q^\nu]\Lambda_b(p)\;,
\end{eqnarray}
with the definitions
\begin{eqnarray}
& &L_1=\frac{1}{2}(f_1+g_1)\;,\;\;\;\; R_1=\frac{1}{2}(f_1-g_1)\;,
\nonumber\\
& &L_2=-\frac{1}{2}(f_2+g_2)\;,\;\;\;\; R_2=-\frac{1}{2}(f_2-g_2)\;.
\end{eqnarray}
Since the form factor $L_1$ dominates for unpolarized $\Lambda_b$ baryon
decays, we shall evaluate only $L_1$ in the present work.

The hadronic matrix element involves both nonperturbative and perturbative
dynamics. To evaluate it, we employ factorization theorems, in which the
former is separated from the latter, and absorbed into universal baryon
wave functions. After performing the factorization, we compute the
perturbative part, {\it i.e.}, the hard subamplitude, order by order
reliably, and convolute it with the baryon wave functions. Certainly, this
factorization picture should fail at some power of $1/M_{\Lambda_b}$,
$M_{\Lambda_b}$ being the $\Lambda_b$ baryon mass, where nonfactorizable
soft gluon exchanges do not cancel exactly. We have derived the
leading-power factorization theorem for heavy meson decays in
\cite{LY1}. This formalism can be extended to heavy baryon decays
straightforwardly, whose basic ideas will be briefly reviewed below.

The lowest-order diagrams for the decay amplitude at the quark level are
exhibited in Fig.~1, where the $b$ quark is represented by a double
line, and the symbol $\times$ denotes the electroweak vertex, from which the
lepton pair emerges. The two gluons are hard, when the proton
recoils fast, which give the necessary momentum transfer to make the
outgoing quarks move collinearly and form the proton. The diagrams
Figs.~1(g)-1(n) were not considered in \cite{RA}, because they were assumed
to be suppressed by $1/M_{\Lambda_b}$. We shall demonstrate that
these diagrams in fact give contributions of the same order as those from
Figs.~1(a)-1(f).

We then consider
radiative corrections to Fig.~1, as shown in Fig.~2, where the bubble
represents the two hard gluons, and explain how to absorb them into the
hard subamplitude and the baryon wave functions. There are three leading
momentum regions for radiative corrections, from which important
contributions arise. These regions are collinear, when the loop momentum is
parallel to an energetic light quark, soft, when the loop momentum is much
smaller than $M_{\Lambda_b}$, and hard, when the loop momentum is of order
$M_{\Lambda_b}$. We associate small transverse
momenta $k_T$ with the valence quarks, which serve as an infrared
cutoff. The collinear and soft contributions may overlap to give double
logarithms. While the hard contribution gives only single logarithms.

Because of the inclusion of the transverse degrees of freedom,
the factorization should be constructed in the impact parameter $b$ space,
with $b$ conjugate to $k_T$ \cite{LY1,CS}. In the
axial gauge $n\cdot A=0$, $n$ being a gauge vector and $A$ the gauge field,
the two-particle reducible corrections on the proton side, like Figs.~2(a)
and 2(b), have the double logarithms $\ln^2(M^2_{\Lambda_b}b^2)$ in the fast
recoil region and the single soft logarithms  $\ln(b\mu)$, $\mu$ being the
renormalization scale. Naturally, these corrections are
absorbed into the proton wave function. The two-particle irreducible
corrections, with the gluon attaching a quark in the $\Lambda_b$ baryon and
a quark in the proton, like Figs.~2(c) and 2(d), contain only the single
logarithms $\ln(b\mu)$, whose effects are less important, and will be
neglected in the following analysis.

On the $\Lambda_b$ baryon side, Fig.~2(e), giving the self-energy correction
to the massive $b$ quark, produces only soft divergences. If the light
valence quarks move slowly, collinear divergences in Figs.~2(f) and 2(g)
will not be pinched \cite{LY1}, and these diagrams also give only soft
divergences. However, there is probability, though small, of finding the
light quarks in the $\Lambda_b$ baryon with longitudinal momenta of order
$M_{\Lambda_b}$. Therefore, Figs.~2(f) and 2(g) may still contribute
collinear divergences. In conclusion, reducible corrections on the
$\Lambda_b$ baryon side are dominated by soft dynamics, but contain weak
double logarithms with the collinear ones suppressed. These corrections are
absorbed into the $\Lambda_b$ baryon wave function. The remaining important
contributions from Fig.~2, with the collinear and soft enhancements
subtracted, are then dominated by short-distance dynamics characterized by
the single logarithms $\ln(M_{\Lambda_b}/\mu)$, and absorbed into the hard
subamplitude.

The kinematics is as follows. The momentum of the $\Lambda_b$ baryon
at rest is $p=(p^+, p^-, {\bf 0})$ with $p^+=p^-=M_{\Lambda_b}/\sqrt{2}$.
The momenta of the valence quarks in the $\Lambda_b$ baryon are
\begin{eqnarray}
k_1=(p^+, x_1 p^-, {\bf k}_{1T})\;,\;\;\;\;
k_2=(0 , x_2 p^-, {\bf k}_{2T})\;,\;\;\;\;
k_3=(0 , x_3 p^-, {\bf k}_{3T})\;,
\end{eqnarray}
where $k_1$ is associated with the $b$ quark, and $x_i$ are the momentum
fractions. The momentum of the proton, which recoils in the plus direction,
is chosen as $p^\prime=({p^\prime}^+, 0,{\bf 0})$ with
$p^{\prime +}=\rho p^+$. The parameter $\rho$, $0\le\rho\le 1$, defined by
\begin{equation}
\rho=\frac{2p\cdot p^\prime}{M_{\Lambda_b}^2}\;,
\end{equation}
is related to the energy fraction of the proton. The invariant mass of the
lepton-neutrino pair is then given by $m_l^2=(1-\rho)M_{\Lambda_b}^2$.
The momenta of the valence quarks in the proton are
\begin{eqnarray}
k_1^\prime=(x_1^\prime p^{\prime +}, 0, {\bf k}_{1T}^\prime)\;,\;\;\;\;
k_2^\prime=(x_2^\prime p^{\prime +}, 0, {\bf k}_{2T}^\prime)\;,\;\;\;\;
k_3^\prime=(x_3^\prime p^{\prime +}, 0, {\bf k}_{3T}^\prime)\;.
\end{eqnarray}
All the transverse momenta are assumed to be much smaller than
$M_{\Lambda_b}$, since the valence quarks are near or on the mass shell.
Otherwise, the quark lines should be absorbed into the hard subamplitude.
In this sense the transverse momenta play the role of a factorization scale,
above which PQCD is reliable, and below which QCD dynamics is parametrized
into the baryon wave functions. For a similar reason, the fraction $x_1$
should be close to unity, as reflected by the behavior of the $\Lambda_b$
baryon wave function in Sect. IV, such that $k_1^2$ is roughly equal to
$M_b^2$.

According to the above explanation, the factorization formula for the form
factor $L_1$ is expressed as
\begin{eqnarray}
L_1 &=& \int_0^1 [dx^\prime][dx]\int [d^2b]
\bar{\Psi}_{P\alpha'\beta'\gamma'}(x_i^\prime,b_i,p^\prime,\mu)
\nonumber\\
& &\times  H^{\alpha'\beta'\gamma'\alpha\beta\gamma}
(x_i^\prime,x_i,b_i,M_{\Lambda_b},\mu)
\Psi_{{\Lambda_b}\alpha\beta\gamma}(x_i,b_i,p,\mu)\;,
\end{eqnarray}
with the symbols
\begin{eqnarray}
[dx] = \prod_{i=1}^3dx_i \delta\left( 1-\sum_{i=1}^3 x_i \right)\;,
\;\;\;\;[ d^2b ] = \prod_{i=1}^2\frac{d^2b_i}{(2\pi)^2}\;,
\end{eqnarray}
where the $\delta$-function is the consequence of the momentum conservation.
The symbol with a prime, $[dx']$, is defined similarly. Note that there are
no the variables $b'_i$. Since we shall drop the transverse momenta carried
by the internal quarks in the hard subamplitude, the transverse separations
among the valence quarks of the $\Lambda_b$ baryon are equal to those
of the proton. It has been shown that this approximation, simplifying the
analysis, gives reasonable predictions for the proton form factor \cite{L}.

The proton distribution amplitude $\Psi_P$ is defined, in the transverse
momentum space, by
\begin{eqnarray}
\Psi_{P\alpha\beta\gamma}&=&\frac{1}{2\sqrt{2}N_c}\int \prod_{l=1}^{2}
\frac{d y_{l}^{-}d{\bf y}_l}
{(2\pi)^{3}}e^{\textstyle ik'_{l}\cdot y_{l}}\epsilon^{abc}
\langle 0|T[u_{\alpha}^{a}(y_{1})u_{\beta}^{b}
(y_{2})d_{\gamma}^{c}(0)]|P\rangle\;,
\nonumber \\
&=&\frac{f_{P}(\mu)}{8\sqrt{2}N_{c}}
[(\,/\llap p'C)_{\alpha\beta}(\gamma_{5}P)_{\gamma}V(k'_{i},p',\mu)
+(\,/\llap p'\gamma_{5}C)_{\alpha\beta}P_{\gamma}A(k'_{i},p',\mu)
\nonumber \\
& &\mbox{ }-(\sigma_{\mu\nu}p^{\prime\nu}C)_{\alpha\beta}(\gamma^{\mu}
\gamma_{5}P)_{\gamma}T(k'_{i},p',\mu)]\; ,
\label{4}
\end{eqnarray}
where $N_{c}=3$ is the color number, $|P\rangle$ the proton state,
$u$ and $d$ the quark fields, $a$, $b$ and $c$ the color indices, and
$\alpha$, $\beta$ and $\gamma$ the spinor indices. The second form
shows the explicit Dirac matrix structure \cite{CZ1} with the normalization
constant $f_{P}$, the proton spinor $P$, the charge conjugation
matrix $C$, and $\sigma_{\mu\nu}\equiv[\gamma_{\mu},\gamma_{\nu}]/2$.
Using the permutation symmetry and the constraint that the
total isospin of the three quarks is equal to $1/2$, the three functions
$V$, $A$, and $T$ are related to a single function $\Psi$ through \cite{CZ1}
\begin{eqnarray}
& &V(k'_1,k'_2,k'_3,p',\mu)=\frac{1}{2}\left[\Psi(k'_2,k'_1,k'_3,p',\mu)
+\Psi(k'_1,k'_2,k'_3,p',\mu)\right]\;,
\nonumber \\
& &A(k'_1,k'_2,k'_3,p',\mu)=\frac{1}{2}\left[\Psi(k'_2,k'_1,k'_3,p',\mu)
-\Psi(k'_1,k'_2,k'_3,p',\mu)\right]\;,
\nonumber \\
& &T(k'_1,k'_2,k'_3,p',\mu)=\frac{1}{2}\left[\Psi(k'_1,k'_3,k'_2,p',\mu)+
\Psi(k'_2,k'_3,k'_1,p',\mu)\right]\;.
\label{u2}
\end{eqnarray}

The structure of the $\Lambda_b$ baryon distribution amplitude
$\Psi_{\Lambda_b}$ is simplified in the heavy quark limit. Under the
assumptions that the spin and orbital degrees of freedom of the light quark
system are decoupled, and the $\Lambda_b$ baryon is in the ground state
($s$-wave), $\Psi_{\Lambda_b}$ is expressed as \cite{RA}
\begin{eqnarray}
\Psi_{\Lambda_b\alpha\beta\gamma}=\frac{f_{\Lambda_b}}{8\sqrt{2}N_c}
[ (\,/\llap p+M_{\Lambda_b})\gamma_5 C ]_{\beta\gamma} 
\Lambda_{b\alpha}\Phi(k_1,k_2,k_3,p,\mu)\;,
\end{eqnarray}
where $f_{\Lambda_b}$ is the normalization constant, and $\Lambda_b$ is
the $\Lambda_b$ baryon spinor. There is also only a single wave function
$\Phi$ associated with the $\Lambda_b$ baryon.
\vskip 1.0cm

\centerline{\large\bf III. SUDAKOV RESUMMATION}
\vskip 0.5cm

The large logarithms, appearing in the wave functions and the hard
subamplitude, should be organized by the resummation technique \cite{CS} and
the renormalization-group (RG) method. The RG summation of the single
logarithms $\ln(M_{\Lambda_b}/\mu)$ and $\ln(b\mu)$ leads to the evolution
from the characteristic scale of the hard subamplitude to the factorization
scale $1/b$. The resummation of the double logarithms
$\ln^2(M_{\Lambda_b}b)$ exhibits Sudakov suppression in the large-$b$
(long-distance) region, such that PQCD analysis becomes relatively reliable
at the energy scale of $M_{\Lambda_b}$. The standard derivation of a
Sudakov factor is referred to \cite{CS,BS}.

The result of the Sudakov resummation for the proton wave function $\Psi_P$
is given by
\begin{eqnarray}
\Psi(x'_i,b_i,p',\mu) &=& \exp\left[-\sum_{l=1}^3
s(cw,x'_l p^{\prime +})-3 \int_{cw}^\mu \frac{d\bar{\mu}}{\bar{\mu}}
\gamma_q(\alpha_s(\bar{\mu}))\right]
\nonumber\\
& &\times \psi(x'_i,cw)\;,
\label{rp}
\end{eqnarray}
with the quark anomalous dimension $\gamma_{q}(\alpha_s)=-\alpha_{s}/\pi$.
The exponent $s$, corresponding to the double-logarithm evolution, is
written as \cite{BS}
\begin{equation}
s(w,Q)=\int_{w}^{Q}\frac{d p}{p}\left[\ln\left(\frac{Q}
{p}\right)A(\alpha_s(p))+B(\alpha_s(p))\right]\;,
\label{fsl}
\end{equation}
where the anomalous dimensions $A$ to two loops and $B$ to one loop are
\begin{eqnarray}
A&=&{\cal C}_F\frac{\alpha_s}{\pi}+\left[\frac{67}{9}-\frac{\pi^2}{3}
-\frac{10}{27}n_f+\frac{8}{3}\beta_0\ln\left(\frac{e^{\gamma_E}}{2}\right)
\right]\left(\frac{\alpha_s}{\pi}\right)^2\;,
\nonumber \\
B&=&\frac{2}{3}\frac{\alpha_s}{\pi}\ln\left(\frac{e^{2\gamma_E-1}}
{2}\right)\;,
\end{eqnarray}
$n_f=4$ for $\Lambda_b$ baryon decays being the flavor number, and
$\gamma_E$ the Euler constant. The two-loop running coupling constant,
\begin{equation}
\frac{\alpha_s(\mu)}{\pi}=\frac{1}{\beta_0\ln(\mu^2/\Lambda^2)}-
\frac{\beta_1}{\beta_0^3}\frac{\ln\ln(\mu^2/\Lambda^2)}
{\ln^2(\mu^2/\Lambda^2)}\;,
\label{ral}
\end{equation}
with the coefficients
\begin{eqnarray}
& &\beta_{0}=\frac{33-2n_{f}}{12}\;,\;\;\;\beta_{1}=\frac{153-19n_{f}}{24}\;,
\label{12}
\end{eqnarray}
and the QCD scale $\Lambda\equiv \Lambda_{\rm QCD}$, will be substituted
into Eq.~(\ref{fsl}). Compared to \cite{RA}, our expression of the
Sudakov factor with the accuracy up to single logarithms is more complete,
since we have used the two-loop running coupling constant in
Eq.~(\ref{ral}). The second exponent in Eq.~(\ref{rp}), {\it i.e.},
the integral containing $\gamma_q$, corresponds to the single-logarithm RG
evolution. The function $\psi$, obtained by factoring the $p^{\prime +}$
dependence out of $\Psi$, coincides with the conventional parton model.

The infrared cutoff $cw$ is the inverse of a typical transverse distance
among the three valence quarks as stated before. The recent PQCD analysis
of the proton form factor shows that a reasonable choice of $w$ is
\cite{JKB}
\begin{eqnarray}
w=\min\left(\frac{1}{b_1}, \frac{1}{b_2}, \frac{1}{b_3}\right)\;,
\label{wp}
\end{eqnarray}
with $b_3=|{\bf b}_1-{\bf b}_2|$. The best fit to the experimental data
of the proton form factor determines the parameter $c=1.14$ \cite{KLS}. The
introduction of this parameter $c$ is natural from the viewpoint of the
resummation, since the scale $cw$, with $c$ of order unity, is as
appropriate as $w$ \cite{CS,BS}. The variation of $c$, with $cw$ being
the factorization scale, represents the different partitions of radiative
corrections into the perturbative Sudakov factor and the nonperturbative
wave function $\psi$. As all $1/b_i$ are much larger than
$\Lambda$, the Sudakov form factor does not give suppression. As one of
these scales gets close to $\Lambda$, the Sudakov factor tends to zero and
suppresses this region. It is easy to observe that choosing the infrared
cutoff in Eq.~(\ref{wp}) suppresses all the infrared divergences.

The Sudakov resummation for a heavy-light system, such as the $B$ meson,
has been developed in \cite{LY1}. The resummation for the $\Lambda_b$
baryon wave function is similar, and the result is
\begin{eqnarray}
\Phi(x_i,b_i,p,\mu) &=&\exp\left[-\sum_{l=2}^3 s(cw,x_lp^-)
-3 \int_{cw}^\mu \frac{d\bar{\mu}}{\bar{\mu}}\gamma_q(\alpha_s(\bar{\mu}))
\right]
\nonumber\\
& &\times\phi(x_i,cw)\;.
\label{rl}
\end{eqnarray}
Note the lack of the exponent $s$ for the $b$ quark. Since the $b$ quark is
heavy, radiative corrections associated with it do not produce double
logarithms. While the coefficient of the anomalous dimension $\gamma_q$ is
still 3, because the RG evolution is determined by the ultraviolet structure
of loop corrections, which are not affected by the heavy quark mass. The
parameter $c$ is also set to 1.14 for convenience.

Working out the contraction of $\bar{\Psi}_{P\alpha'\beta'\gamma'}
H^{\alpha'\beta'\gamma'\alpha\beta\gamma}
\Psi_{{\Lambda_b}\alpha\beta\gamma}$ in the momentum space, whose explicit
expression from each quark-level decay diagram in Fig.~1 is listed in
Table I, we extract the hard subamplitude $H$. To simplify the formalism, we
have applied the approximation $M_b\approx M_{\Lambda_b}$, and neglected
the transverse momentum dependence of the internal quark propagators as
mentioned before. The RG evolution of $H$ in the $b$ space is written as
\begin{eqnarray}
H(x_{i}',x_{i},b_i,M_{\Lambda_b},\mu)&=&
\exp\left[-3\sum_{l=1}^2\int^{t_{l}}_{\mu}\frac{d\bar{\mu}}{\bar{\mu}}\,
\gamma_{q}\left(\alpha_s(\bar{\mu})\right)\right]
\nonumber \\
& &\times H(x_{i}',x_{i},b_i,M_{\Lambda_b},t_1,t_2)\;,
\label{9}
\end{eqnarray}
where the hard scales $t_l$ will be specified below. The two arguments
$t_{1}$ and $t_{2}$ of $H$ imply that each running coupling constant
$\alpha_s$ is evaluated at the mass scale of the corresponding hard gluon.

\vskip 1.0cm

\centerline{\large\bf IV. FACTORIZATION FORMULA}
\vskip 0.5cm

Employing a series of permutations of the valence quark kinematic variables,
the summation of the expressions in Table I reduces to two terms.
Combining Eqs.~(\ref{rp}), (\ref{rl}), and (\ref{9}), we obtain
the factorization formula,
\begin{eqnarray}
L_1 &=& \frac{2\pi}{27}\int_0^1 [dx^\prime][dx]
\int_0^{1/\Lambda} b_1db_1 b_2db_2\int_0^{2\pi} d\theta
f_P(cw)f_{\Lambda_b}
\nonumber\\
& &\times \sum_{j=1}^2 H_j(x'_i,x_i,b_i,M_{\Lambda_b},t_{jl})
\Psi_j(x_i^\prime,x_i,cw)
\nonumber\\
& &\times \exp[-S(x'_i,x_i,cw,M_{\Lambda_b},t_{jl})]\;,
\label{l1}
\end{eqnarray}
where the variable $\theta$ is the angle between ${\bf b}_1$ and
${\bf b}_2$. The expressions of $H_j$ are
\begin{eqnarray}
& &H_1=\alpha_{s}(t_{11})\alpha_{s}(t_{12})
K_{0}(\sqrt{(1-x_{1})(1-x_{1}')\rho}M_{\Lambda_b} b_1)
K_{0}(\sqrt{x_{2}x_{2}'\rho}M_{\Lambda_b} b_2),
\nonumber \\
& &H_2=\alpha_{s}(t_{21})\alpha_{s}(t_{22})
K_{0}(\sqrt{x_{1}x_{1}'\rho}M_{\Lambda_b} b_1)
K_{0}(\sqrt{x_{2}x_{2}'\rho}M_{\Lambda_b} b_2)\;,
\label{k}
\end{eqnarray}
with $K_{0}$ the modified Bessel function of order zero. The functions
$\Psi_j$, which group together the products of the initial and final wave
functions, are, in terms of the notations,
\begin{eqnarray}
\phi_{123}\equiv \phi(x_1,x_2,x_3)\;,\;\;\;\;
\psi_{123}\equiv \psi(x_1',x_2',x_3',cw)\;,
\end{eqnarray}
given by
\begin{eqnarray}
\Psi_1 &=& 
\frac{2(\psi_{312}\phi_{132}+2T_{132}\phi_{132}+\psi_{213}\phi_{123}
+2T_{123}\phi_{123})}{(1-x_1)(1-x'_1)}
\nonumber \\ & &
+\frac{x'_1 (\psi_{312}\phi_{132}+2T_{132}\phi_{132}
+\psi_{213}\phi_{123}+2T_{123}\phi_{123})}{(1-x'_1)^2\rho} \;,
\label{psi1}\\
\Psi_2 &=& 
 \frac{2T_{312}\phi_{312}}{(1-x_3)(1-x'_1)}
+\frac{-\psi_{132}\phi_{312}}{(1-x_3)(1-x'_2)}
+\frac{2x_2(\rho-1)T_{312}\phi_{312}}{(1-x'_1)\rho[(1-x_2)x'_1\rho+x_2]}
\nonumber \\ & &
+\frac{ [(1-x'_2-x'_3 x_1)\rho-x_1] \psi_{132}\phi_{312}}
      {(1-x'_3)(1-x_1)\rho(x'_2\rho-1)}
\nonumber\\ & & 
+2\frac{[(1-2x_1+x'_2-x'_2\rho)\rho+x_1] T_{312}\phi_{312}}
      {(1-x'_3)(1-x_1)\rho(x'_2\rho-1)}
\nonumber \\ & & 
+\frac{ [(1-2x_2+x'_1-x'_1\rho)\rho+x_2]\psi_{132}\phi_{312}}
      {(1-x'_3)(1-x_2)\rho(x'_1\rho-1)}
\nonumber\\ & &
+2\frac{[(1-x_2-x'_1-(1-x_2)(1-x'_3)\rho)\rho-x_2]T_{312}\phi_{312}}
      {(1-x'_3)(1-x_2)\rho(x'_1\rho-1)}\;.
\label{psi2}
\end{eqnarray}
The functions $T$ are related to $\psi$ in a similar way to Eq.~(\ref{u2}).
The complete Sudakov exponent $S$ is written as
\begin{eqnarray}
S&=&\sum_{l=2}^3 s(cw,x_lp^-)+
3\int_{cw}^{t_{j1}}\frac{d\bar{\mu}}{\bar{\mu}}
\gamma_{q}\left(\alpha_s(\bar{\mu})\right)
\nonumber \\
& &+\sum_{l=1}^3 s(cw,x'_lp^{\prime +})+
3\int_{cw}^{t_{j2}}\frac{d\bar{\mu}}{\bar{\mu}}
\gamma_{q}\left(\alpha_s(\bar{\mu})\right)\;,
\end{eqnarray}
with the hard scales 
\begin{eqnarray}
& &t_{11}=\max\left[\sqrt{(1-x_{1})(1-x_{1}')\rho}M_{\Lambda_b},
1/b_1,cw\right]\;,
\nonumber \\
& &t_{21}=\max\left[\sqrt{x_{1}x_{1}'\rho}M_{\Lambda_b}, 1/b_1,cw\right]\;,
\nonumber \\
& &t_{12}=t_{22}=\max\left[\sqrt{x_{2}x_{2}'\rho}M_{\Lambda_b},
1/b_2,cw\right]\; .
\label{tt}
\end{eqnarray}
The first scales in the brackets are associated with the longitudinal
momenta of the exchanged gluons and the second ones with the transverse
momenta. All $t_{jl}$ should be larger than the factorization scale $cw$
by definition as shown above.

For the wave function $\psi$, we adopt the KS model \cite{KS}, which has
been found to give predictions for the proton form factor in better
agreement with the data compared to the Chernyak-Zhitnitsky model
\cite{CZ1}. The KS model is decomposed as \cite{LB1,BL}
\begin{equation}
\psi(x_{i},w)=\psi_{as}(x_{i})\sum_{j=0}^{5}N_{j}
\left[\frac{\alpha_{s}(w)}{\alpha_{s}(\mu_{0})}\right]^{b_{j}/(4\beta_0)}
a_{j}A_{j}(x_{i})\; ,
\label{wf}
\end{equation}
with $\mu_{0}\approx 1$ GeV and $\psi_{as}=120x_{1}x_{2}x_{3}$. The
constants $N_{j}$, $a_{j}$ and $b_{j}$, and the Appel polynomials $A_{j}$
are listed in Table II. The evolution of the normalization constant $f_{P}$
is given by \cite{CZ1}
\begin{equation}
f_{P}(w)=f_{P}(\mu_{0})\left[\frac{\alpha_{s}(w)}{\alpha_{s}(\mu_{0})}
\right]^{1/(6\beta_0)}\; ,
\label{fn}
\end{equation}
with $f_{P}(\mu_{0})=(5.2\pm 0.3)\times 10^{-3}$ GeV$^2$.

We do not consider the evolution of the $\Lambda_b$ wave function $\phi$ in
the scale $cw$, because it is still not known yet. Similarly, the running of
the normalization constant $f_{\Lambda_b}$ is also neglected, which is
assumed to be $f_{\Lambda_b}=f_P(\mu_0)$ for simplicity \cite{RA}. It is
convenient to use the new dimensionless variables,
\begin{eqnarray}
\zeta = \frac{x_2}{x_2+x_3}\;,\;\;\;\;
\eta = x_2+x_3 \;,
\end{eqnarray}
when presenting the model for the $\Lambda_b$ baryon wave function $\phi$.
We choose \cite{S}
\begin{eqnarray}
\phi(\zeta,\eta)= N \eta^2\zeta (1-\eta)(1-\zeta)
\exp\left[-\frac{M_{\Lambda_b}^2}{2\beta^2(1-\eta)}
-\frac{m_q^2}{2\beta^2\eta\zeta (1-\zeta)}\right],
\end{eqnarray}
for $\phi_{123}$ and $\phi_{132}$, where the parameter $\beta$ can be fixed 
by experimental data of $\Lambda_b$ baryon decays, and $m_q$ is the mass
of the light degrees of freedom. Accordingly, we have
\begin{eqnarray}
\phi(\zeta,\eta)&=&N \eta^2\zeta (1-\eta)(1-\zeta)
\nonumber\\
& &\times\exp\left[-\frac{M_{\Lambda_b}^2}{2\beta^2\eta(1-\zeta)}
-\frac{m_q^2(1-\eta+\eta\zeta)}
{2\beta^2\eta\zeta (1-\eta)}\right]\;,
\end{eqnarray}
for $\phi_{312}$. The normalization constant $N$ is determined by
\begin{equation}
\int_0^1 d\zeta \int_0^1 \eta d\eta\phi(\zeta,\eta)=1\;.
\end{equation}
In the next section we shall investigate how the predictions for the form
factor $L_1$ vary with the parameter $\beta$.

\vskip 1.0cm

\centerline{\large\bf V. NUMERICAL RESULTS}
\vskip 0.5cm
We are now ready to compute the form factor $L_1$ in Eq.~(\ref{l1})
numerically, adopting $M_{\Lambda_b}=5.621$ GeV, $\Lambda=0.2$ GeV and
$m_q=0.8$ GeV. We consider two typical values of the parameter $\beta$, 1.0
and 2.0, and the corresponding dependences of $L_1$ on the cutoff of $b_1$
and $b_2$, $b_{1c}=b_{2c}=b_c$, for the proton energy fraction $\rho=0.8$,
0.9, and 1.0, are displayed in Figs.~3 and 4, respectively. It is observed
that the rise of all the curves indeed saturates at about
$b_c=0.85/\Lambda$, indicating that the large-$b$ region does not
contribute because of Sudakov suppression. Figure 3 shows that for
$\rho=0.8$, approximately 47\% of the contribution to $L_1$ comes from the
region with $\alpha_s(1/b_c)/\pi< 0.5$, or equivalently, $b_c< 0.61\Lambda$.
For $\rho=0.9$, 50\% of the contribution is accumulated in this region. As
$\rho=1.0$, the small-$b$ contribution reaches 52\%.

For the larger $\beta=2.0$, which corresponds to a broader $\Lambda_b$
baryon wave function $\phi$, the percentages of perturbative contributions
for $\rho=0.8$, 0.9, and 1.0 increase a bit up to 49\%, 52\%, and 55\%,
respectively, as shown in Fig.~4. Since the long-distance contribution is
less enhanced by $\phi$, the magnitudes of $L_1$ are reduced to about 63\%
of those for $\beta=1.0$. If regarding the region with $\alpha_s/\pi<0.5$ as
being perturbative \cite{LY1}, the PQCD analysis of
the decay $\Lambda_b\to p l \bar\nu$ for $\rho > 0.8 $ is relatively
self-consistent, since the short-distance contributions begin to dominate.
The above percentages of perturbative contributions are smaller than those
in the heavy meson decay $B\to\pi l\nu$, which are roughly 60-70\% near the
high end of the pion energy spectrum \cite{LY1}. This is expected, because
there are more partons to share the baryon momentum, such that the partons
are softer, and Sudakov suppression is weaker.

Compared to the analysis in \cite{RA}, we have taken into account the
resummation effect from the $\Lambda_b$ baryon and the contributions
to the hard subamplitude from Figs.~1(g)-1(n). Recalculating $L_1$ for
$\beta=1.0$ and $\rho=1.0$ without the Sudakov factor associated with the
$\Lambda_b$ baryon, we observe a 37\% decrease. Note that the inclusion of
the Sudakov factor from the heavy baryon causes an enhancement, instead of
a suppression. It is not difficult to understand this result. Since $\Psi_j$
in Eqs.~(\ref{psi1}) and (\ref{psi2}) are not positive definite in the whole
kinematic region, it is possible that the modulation of this Sudakov factor
suppresses the negative portion of the integrand in Eq.~(\ref{l1}) more than
the positive portion.

The more essential improvement
is the consideration of Figs.~1(g)-1(n) in the evaluation of the hard
subamplitude. It was argued that the contributions from these diagrams,
with at least one of the exchanged gluons attaching the $b$ quark, are
suppressed by the power $1/M_{\Lambda_b}$, and thus negligible \cite{RA}.
However, this argument is correct only when the exchanged gluons are soft,
such that the $b$ quark propagators in Figs.~1(g)-1(n) provide a power
suppression by $1/M_{\Lambda_b}$. Before investigating the validity of this
argument, it has contradicted the assumption of hard gluon exchangs, based
on which the PQCD formalism was developed. Our analysis indicates that with
the help of Sudakov suppression, the exchanged gluons are indeed hard
enough, and that Figs.~1(g)-1(n) lead to contributions of the same order as
those from Figs~1(a)-1(f). If considering only the first six diagrams in
Fig.~1 as in \cite{RA}, the result of $L_1$ for $\beta=1.0$ and $\rho=1.0$
reduces by a factor 3. That is, the diagrams ignored in \cite{RA} are
in fact even more important.

At last, we present the variations of $L_1$ and of the differential decay
rate
\begin{eqnarray}
\frac{d \Gamma}{d \rho}&=&|V_{ub}|^2
\frac{G_F^2M_{\Lambda_b}^5}{96\pi^3}\rho^2(3-2\rho)|L_1|^2\;,
\end{eqnarray}
for massless leptons with the proton energy fraction $\rho$. Employing the
Fermi coupling constant $G_F=1.16639\times 10^{-5}$ GeV$^{-2}$ and the CKM
matrix element $V_{ub}=0.003$, results for $\beta=1.0$, 1.5, and 2.0 are
displayed in Figs.~5 and 6, respectively. The curves rise quickly as
$\rho<0.6$, implying the dominance of nonperturbative contributions. Our
predictions become insensitive to the parameter $\beta$ of the $\Lambda_b$
baryon wave function as $\rho>0.8$. Therefore, we conclude that the values
of $L_1$ and of $d\Gamma/d\rho$ for $\rho=1.0$ are about
$L_1=2.3 \times 10^{-3}$ and $d\Gamma/d\rho=9.4 \times 10^{-21}$ GeV.

\vskip 1.0cm

\centerline{\large \bf VI. CONCLUSION}
\vskip 0.5cm

In this paper we have studied the semileptonic decay
$\Lambda_b\to p l\bar\nu$ as an initial application of the PQCD
factorization theorem to heavy baryon decays. It is found that the
perturbative analysis is reliable near the high end ($\rho> 0.8$) of
the proton energy spectrum. Compared to the previous study \cite{RA}, we
have taken into account the Sudakov resummation for a heavy-light
system (the $\Lambda_b$ baryon) and a complete hard subamplitude.
The quark-level decay diagrams neglected in \cite{RA} turn out to be
very important. Our approach will be applied to the semileptonic
decay $\Lambda_b\to \Lambda_c l\bar\nu$, for which experimental data are
available. We shall try to determine the parameter $\beta$ of the
$\Lambda_b$ baryon wave function from this decay, and then extend the PQCD
formalism to nonleptonic $\Lambda_b$ baryon decays, which are more
challenging. These subjects will be published elsewhere.

\vskip 1.0cm

This work was supported by the National Science Council of Republic of
China under Grant Nos. NSC-88-2112-M-001-041 and NSC-88-2112-M-006-013.

\newpage

Table 1.
The expressions of $\bar{\Psi}_{P\alpha'\beta'\gamma'}
H^{\alpha'\beta'\gamma'\alpha\beta\gamma}
\Psi_{{\Lambda_b}\alpha\beta\gamma}$ from the diagrams in Fig.~1 in terms
of
\begin{eqnarray}
{\cal D}_{ii} &=& 
x_ix'_i\rho M_{\Lambda_b}^2+(k_{iT}-k^\prime_{iT})^2 \;,
\nonumber\\
\bar{\cal{D}}_{ii} &=& 
(1-x_i)(1-x'_i)\rho M_{\Lambda_b}^2+(k_{iT}-k^\prime_{iT})^2 \;.
\nonumber
\end{eqnarray}
For each diagram, (a)-(n), row (1) is the original expression, row (2) is
the series of variable changes necessary for bringing row (1) into the
desired form, and row (3) is the expression after variable changes.

\vskip 1.0cm
\begin{tabular}{rc}
\hline 
Diagram & $\bar{\Psi}_PH\Psi_{\Lambda_b}/(4\pi^2\alpha_s^2
f_Pf_{\Lambda_b}/27)$ \\
\hline 
(a) (1) & 
$\frac{[\psi_{213}\phi_{123}+2T_{123}\phi_{123}]}{(1-x_1)(1-x'_1)
\bar{\cal{D}}_{11}{\cal D}_{33}}$
\\
    (2) &  
$ 2 \leftrightarrow 3$
\\
    (3) &  
$\frac{[\psi_{312}\phi_{132}+2T_{132}\phi_{132}]}{(1-x_1)(1-x'_1)
           \bar{\cal{D}}_{11}{\cal D}_{22}}$
\\
(b) (1) &  0
\\
    (2) &  0
\\
    (3) &  0
\\
(c) (1) & 
$\frac{2T_{123}\phi_{123}}{(1-x_1)(1-x'_2){\cal D}_{22}{\cal D}_{33}}$
\\
    (2) &  
$1 \to 3, 2 \to 1, 3 \to 2$
\\
    (3) & 
$\frac{2T_{312}\phi_{312}}{(1-x_3)(1-x'_1){\cal D}_{11}{\cal D}_{22}}$
\\
(d) (1) &
$\frac{-\psi_{213}\phi_{123}}{(1-x_1)(1-x'_3){\cal D}_{22}{\cal D}_{33}}$
\\
    (2) &
$1 \to 3, 2 \to 1, 3 \to 2$
\\
    (3) &
$\frac{-\psi_{132}\phi_{312}}{(1-x_3)(1-x'_2){\cal D}_{11}{\cal D}_{22}}$
\\
(e) (1) & 0
\\
    (2) & 0
\\
    (3) & 0
\\
(f) (1) & 
$\frac{ [\psi_{213}\phi_{123}+2T_{123}\phi_{123}]}{(1-x_1)(1-x'_1)
\bar{\cal{D}}_{11}{\cal D}_{22}}$
\\
    (2) &
None
\\
    (3) &
$\frac{ [\psi_{213}\phi_{123}+2T_{123}\phi_{123}]}{(1-x_1)(1-x'_1)
\bar{\cal{D}}_{11}{\cal D}_{22}}$
\\
(g) (1) &
$\frac{2x_3(\rho-1) T_{123}\phi_{123}}{(1-x'_2)\rho[(1-x_3)x'_2\rho+x_3]
{\cal D}_{22}{\cal D}_{33}}$
\\
    (2) &
$1 \to 3, 2 \to 1, 3 \to 2$
\\
    (3) &
$\frac{2x_2(\rho-1) T_{312}\phi_{312}}{(1-x'_1)\rho[(1-x_2)x'_1\rho+x_2]
{\cal D}_{11}{\cal D}_{22}}$\\
\hline
\end{tabular}

\newpage

\begin{tabular}{rc}
\hline 
Diagram & $\bar{\Psi}_PH\Psi_{\Lambda_b}/(4\pi^2\alpha_s^2
f_Pf_{\Lambda_b}/27)$ \\
\hline 
(h) (1) & 0
\\
    (2) & 0
\\
    (3) & 0
\\
(i) (1) &
$\frac{x'_1 [\psi_{213}\phi_{123}+2T_{123}\phi_{123}]}{(1-x'_1)^2\rho
\bar{\cal{D}}_{11}{\cal D}_{22}}$
\\
    (2) & None
\\
    (3) &
$\frac{x'_1 [\psi_{213}\phi_{123}+2T_{123}\phi_{123}]}{(1-x'_1)^2\rho
\bar{\cal{D}}_{11}{\cal D}_{22}}$
\\
(j) (1) &
$\frac{ [\psi_{213}\phi_{123}+2T_{123}\phi_{123}]}{(1-x_1)(1-x'_1)
\bar{\cal{D}}_{11}{\cal D}_{22}}$
\\
    (2) & None
\\
    (3) &
$\frac{ [\psi_{213}\phi_{123}+2T_{123}\phi_{123}]}{(1-x_1)(1-x'_1)
\bar{\cal{D}}_{11}{\cal D}_{22}}$
\\
(k) (1) &
$\frac{ [(1-x'_3-x'_1 x_2)\rho-x_2] \psi_{213}\phi_{123}}
{(1-x'_1)(1-x_2)\rho(x'_3\rho-1){\cal D}_{22}{\cal D}_{33}}$
\\      &
$+\frac{ 2[(1-2x_2+x'_3-x'_3\rho)\rho+x_2] T_{123}\phi_{123}}
{(1-x'_1)(1-x_2)\rho(x'_3\rho-1){\cal D}_{22}{\cal D}_{33}}$
\\
    (2) &
$1 \to 3, 2 \to 1, 3 \to 2$
\\
    (3) &
$\frac{ [(1-x'_2-x'_3 x_1)\rho-x_1] \psi_{132}\phi_{312}}
{(1-x'_3)(1-x_1)\rho(x'_2\rho-1){\cal D}_{11}{\cal D}_{22}}$
\\      &
$+\frac{ 2[(1-2x_1+x'_2-x'_2\rho)\rho+x_1] T_{312}\phi_{312}}
{(1-x'_3)(1-x_1)\rho(x'_2\rho-1){\cal D}_{11}{\cal D}_{22}}$
\\
(l) (1) &
$\frac{
[(1-2x_3+x'_2-x'_2\rho)\rho+x_3] 
\psi_{213}\phi_{123}}
{(1-x'_1)(1-x_3)\rho(x'_2\rho-1){\cal D}_{22}{\cal D}_{33}}$
\\       &
$+\frac{2[(1-x_3-x'_2-(1-x_3)(1-x'_1)\rho)\rho-x_3] T_{123}\phi_{123}}
{(1-x'_1)(1-x_3)\rho(x'_2\rho-1){\cal D}_{22}{\cal D}_{33}}$
\\
    (2) &
$1 \to 3, 2 \to 1, 3 \to 2$
\\
    (3) &
$\frac{ [(1-2x_2+x'_1-x'_1\rho)\rho+x_2] \psi_{132}\phi_{312}}
{(1-x'_3)(1-x_2)\rho(x'_1\rho-1){\cal D}_{11}{\cal D}_{22}}$
\\      &
$+\frac{ 2[(1-x_2-x'_1-(1-x_2)(1-x'_3)\rho)\rho-x_2] T_{312}\phi_{312}}
{(1-x'_3)(1-x_2)\rho(x'_1\rho-1){\cal D}_{11}{\cal D}_{22}}$
\\
(m) (1) &
$\frac{ [\psi_{213}\phi_{123}+2T_{123}\phi_{123}]}{(1-x_1)(1-x'_1)
\bar{\cal{D}}_{11}{\cal D}_{33}}$
\\
    (2) &
$ 2 \leftrightarrow 3$
\\
    (3) &
$\frac{ [\psi_{312}\phi_{132}+2T_{132}\phi_{132}]}{(1-x_1)(1-x'_1)
\bar{\cal{D}}_{11}{\cal D}_{22}}$
\\
(n) (1) &
 $\frac{x'_1 [\psi_{213}\phi_{123}+2T_{123}\phi_{123}]}{(1-x'_1)^2\rho
\bar{\cal{D}}_{11}{\cal D}_{33}}$
\\
    (2) & 
$ 2 \leftrightarrow 3$
\\
    (3) & 
$\frac{x'_1 [\psi_{312}\phi_{132}+2T_{132}\phi_{132}]}{(1-x'_1)^2\rho
\bar{\cal{D}}_{11}{\cal D}_{22}}$\\
\hline
\end{tabular}

\newpage

Table 2.
The constants and the Appel polynomials involved in the KS
proton wave function $\psi$ in Eq.~(\ref{wf}) \cite{JSL}.

\[ \begin{array}{clccl} \hline
j   & a_j     & N_{j} & b_{j} & A_{j}(x_{i}) \\ \hline
0   &1.00     & 1     & 0     & 1  \\
1   &0.310    & 21/2  & 20/9  & x_{1}-x_{3} \\
2   &-0.370   & 7/2   & 24/9  & 2-3(x_{1}+x_{3}) \\
3   &0.630    & 63/10 & 32/9  & 2-7(x_{1}+x_{3})+8(x_{1}^{2}
                                          +x_{3}^{2})+4x_{1}x_{3} \\
4   &0.00333  & 567/2 & 40/9  & x_{1}-x_{3}-(4/3)(x_{1}^{2}-x_{3}^{2}) \\
5   &0.0632   & 81/5  & 42/9  & 2-7(x_{1}+x_{3})+14x_{1}x_{3}  \\
    &         &       &       &  +(14/3)(x_{1}^{2}+ x_{3}^{2})\\
\hline
\end{array}  \]

\newpage

\centerline{\large\bf FIGURE CAPTIONS}
\vskip 0.5cm

\noindent
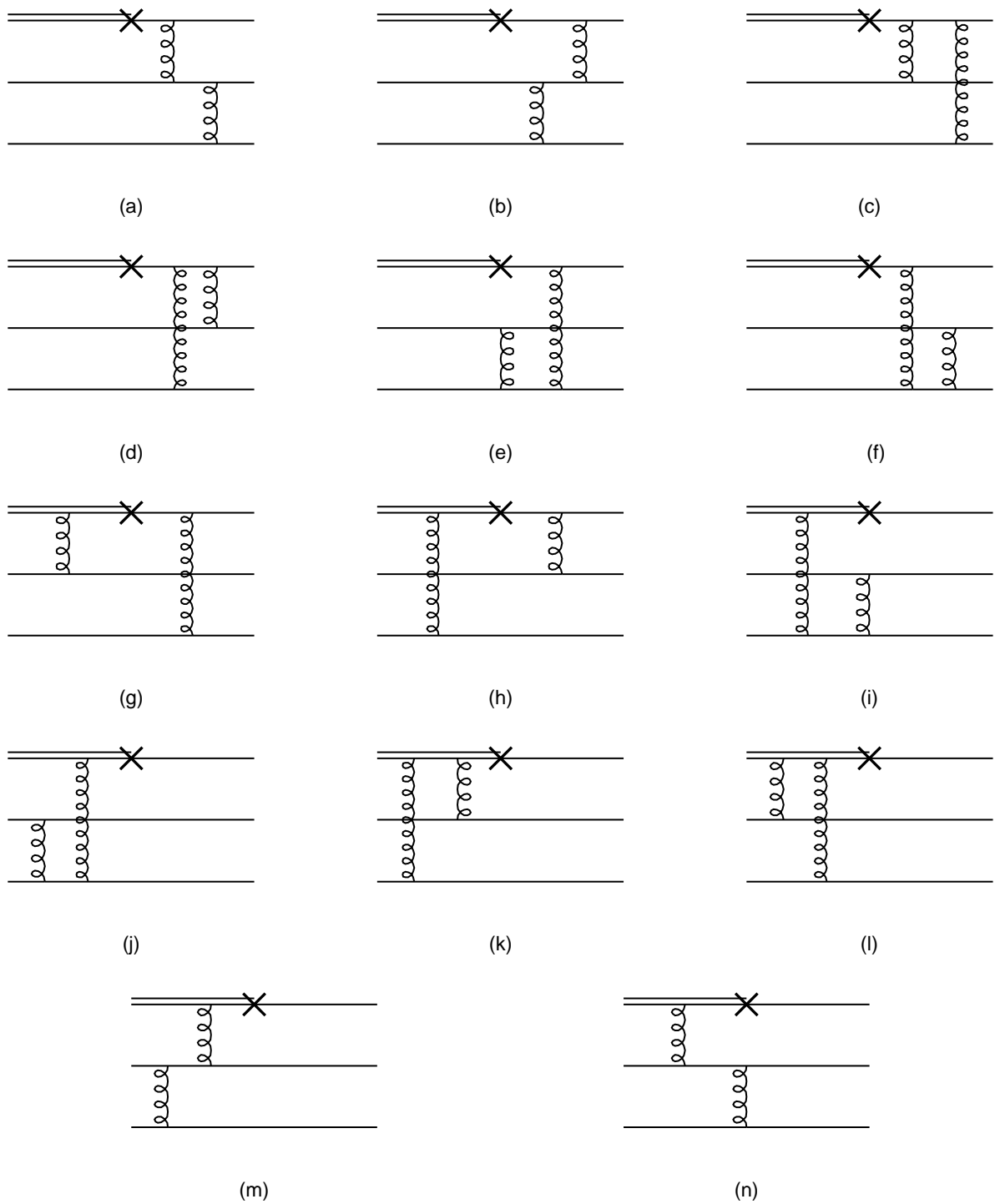
{\bf FIG. 1} Lowest order diagrams for the $\Lambda_b\to pl\bar \nu$
decay.
\vskip 0.3cm

\noindent
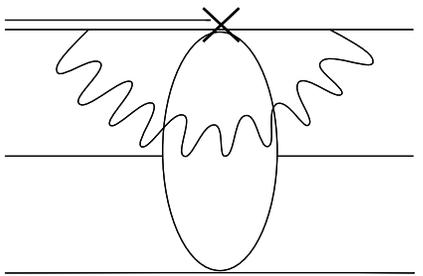
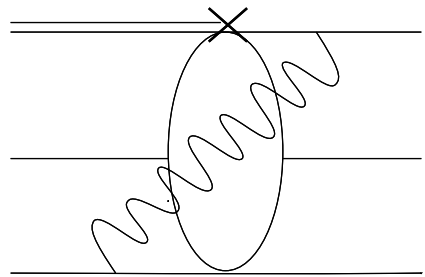
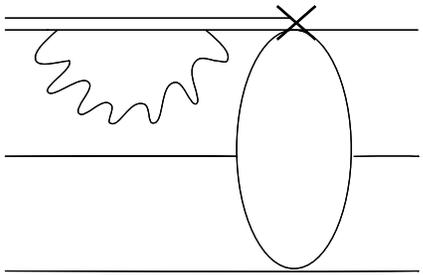
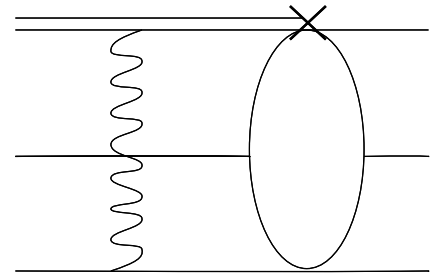
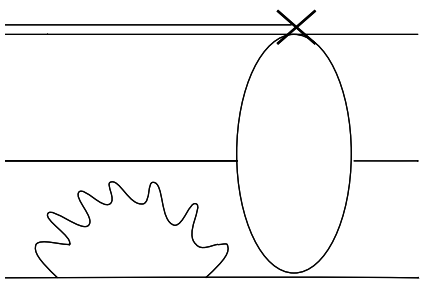
{\bf FIG. 2} Radiative corrections to Fig. 1.
\vskip 0.3cm

\noindent
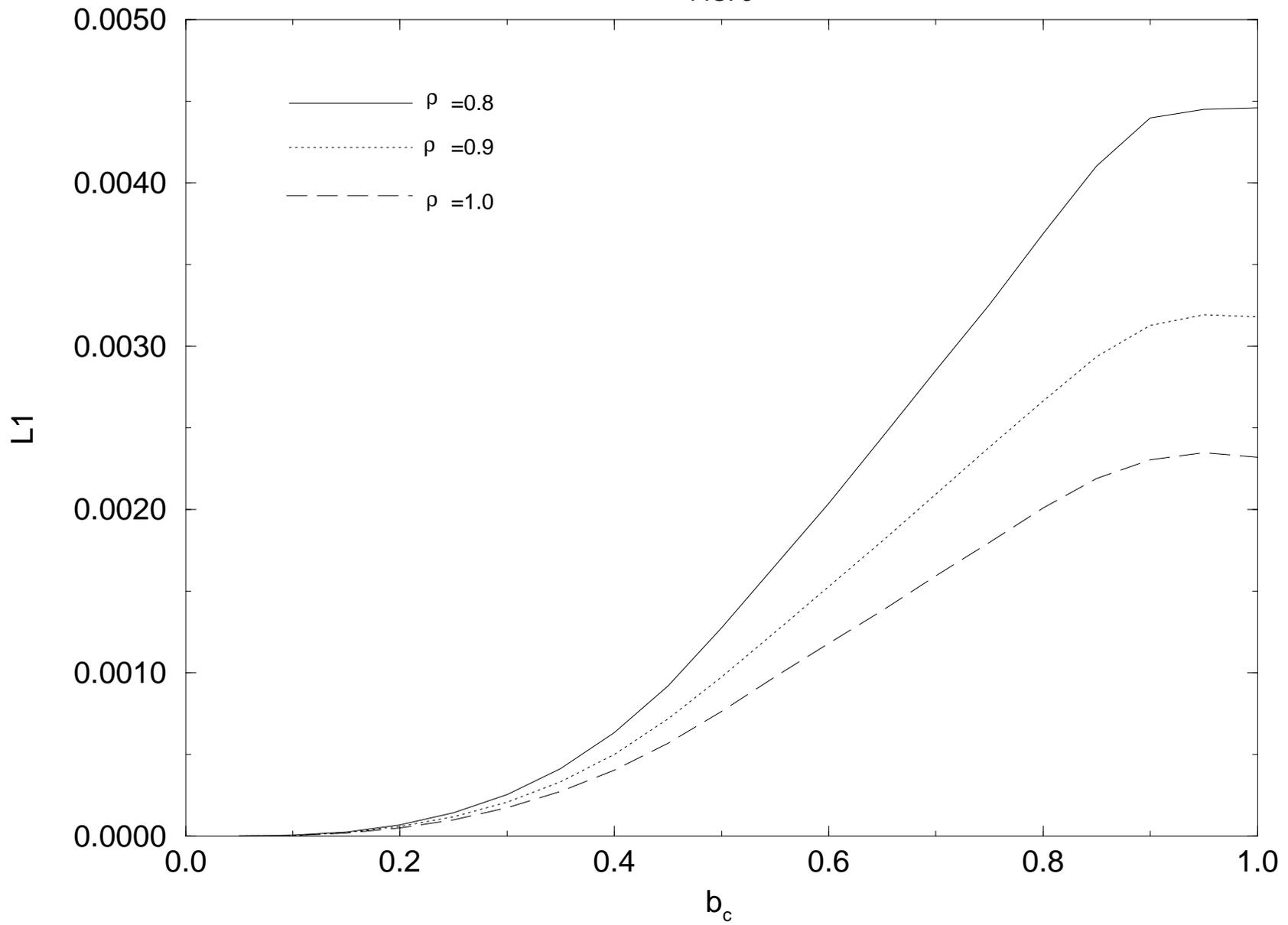
{\bf FIG. 3} Dependence of $L_1$ on $b_c$ for $\beta=1.0$.
             $b_c$ is in unit of $1/\Lambda_{\rm QCD}$. 
\vskip 0.3cm

\noindent
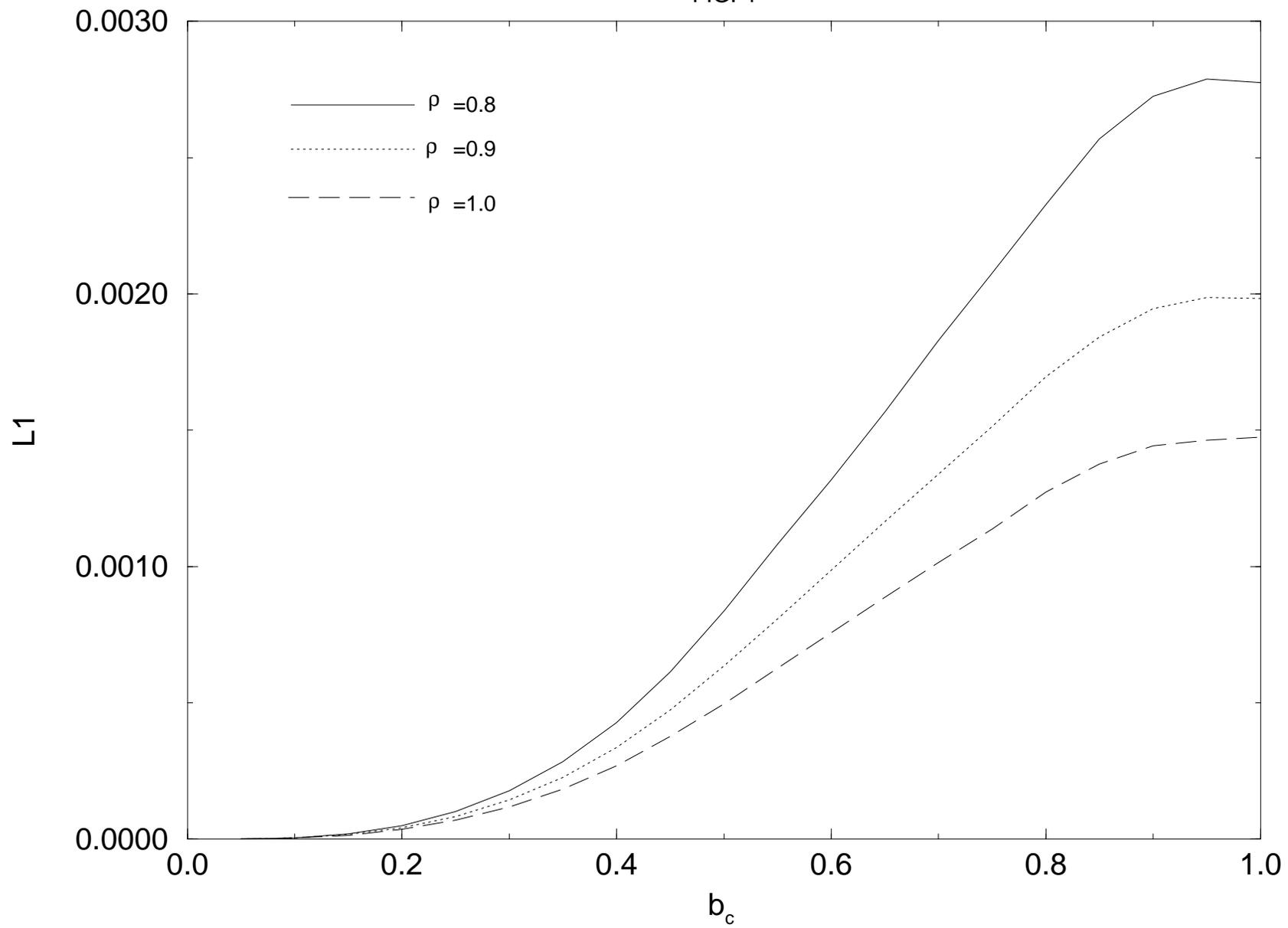
{\bf FIG. 4} Dependence of $L_1$ on $b_c$ for $\beta=2.0$.
             $b_c$ is in unit of $1/\Lambda_{\rm QCD}$. 
\vskip 0.3cm

\noindent
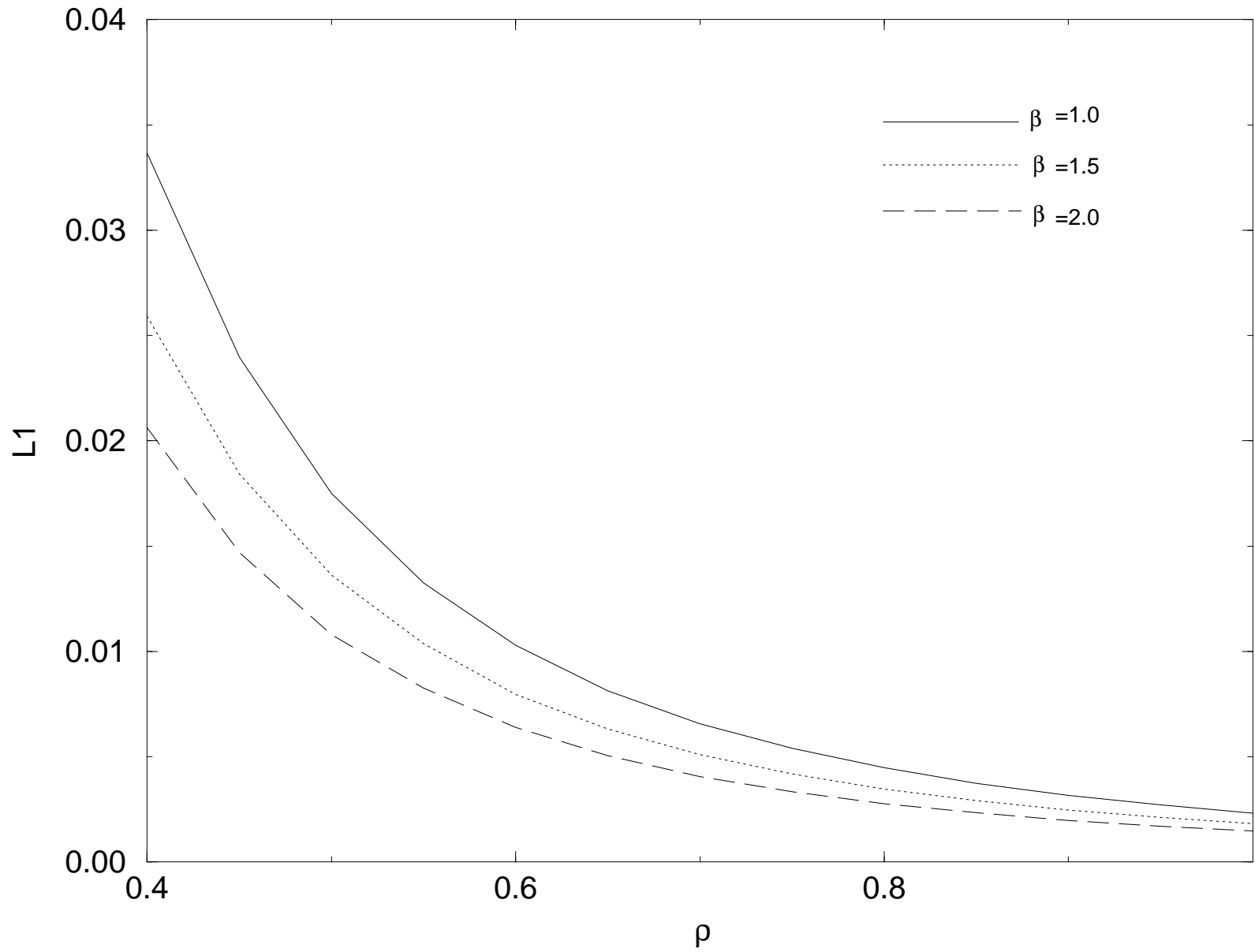
{\bf FIG. 5} Dependence of $L_1$ on $\rho$.
\vskip 0.3cm

\noindent
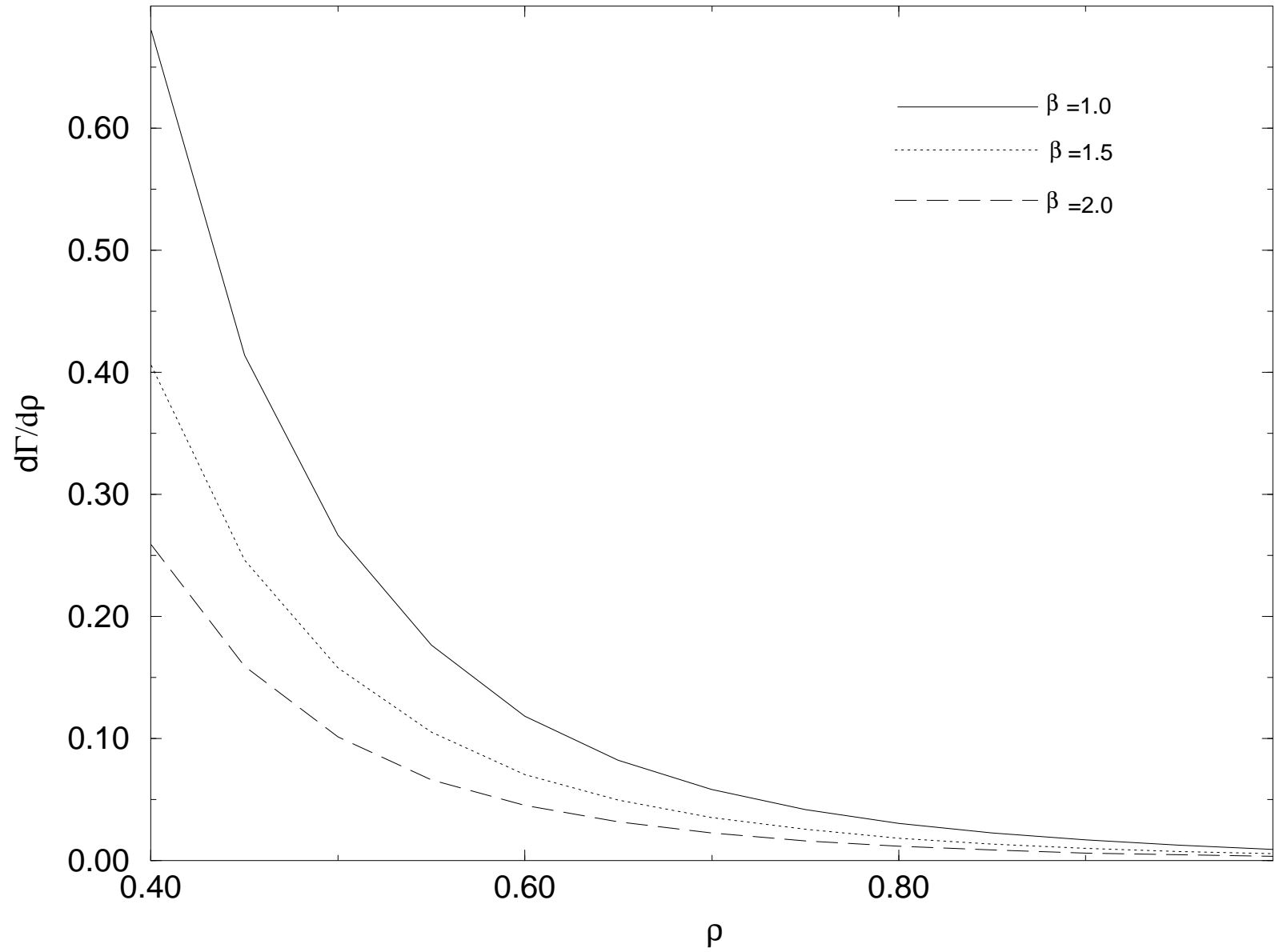
{\bf FIG. 6} Dependence of $d\Gamma/d\rho$ on $\rho$ in unit of
$10^{-18}$ GeV.
\vskip 0.3cm
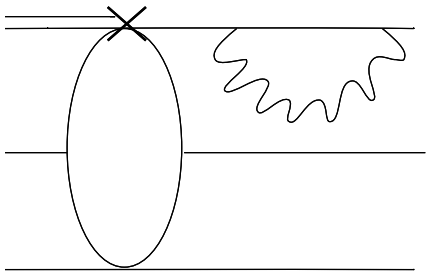
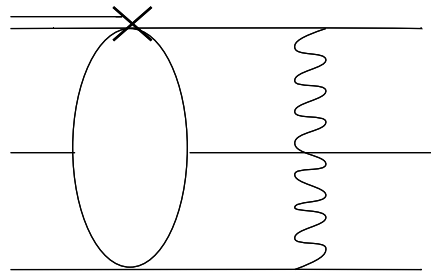
\end{document}